\documentclass[english]{elsarticle}
\usepackage[T1]{fontenc}
\usepackage[latin9]{inputenc}
\usepackage{babel}
\usepackage{amsmath}
\usepackage{amssymb}
\usepackage{graphicx}
\usepackage[unicode=true]
 {hyperref}
\usepackage{breakurl}

\makeatletter

\providecommand{\tabularnewline}{\\}
\newcommand{\lyxdot}{.}

\usepackage {lineno}

\makeatother

\begin{document}

\begin{frontmatter}{\title{High Precision Momentum Calibration of the Magnetic Spectrometers at MAMI for Hypernuclear Binding Energy Determination}
\author[yer]{A. Margaryan\corref{c1}}
\ead{mat@mail.yerphi.am}
\author[gla]{J.R.M. Annand\corref{c2}}
\ead{john.annand@glasgow.ac.uk}
\author[mz]{P. Achenbach}
\author[yer]{R. Ajvazyan}
\author[yer]{H.  Elbakyan}
\author[gla]{R. Montgomery}
\author[send]{S. N. Nakamura}
\author[mz]{J. Pochodzalla}
\author[mz]{F. Schulz}
\author[send]{Y. Toyama}
\author[yer]{S. Zhamkochyan}

\address[yer]{A.I. Alikhanyan National Science Laboratory, Yerevan 0036, Armenia.}

\address[mz]{Institut fur Kernphysik, Johannes Gutenberg Universit\"at, 55099 Mainz, Germany.}

\address[send]{Department of Physics, Tohoku University, Sendai 980-8578, Japan.}

\address[gla]{School of Physics and Astronomy, University of Glasgow, G12 8QQ, Scotland, UK.}

\cortext[c1]{Corresponding author 1}
\cortext[c2]{Corresponding author 2}

\begin{keyword}
Magnetic spectrometers, Cherenkov detectors, Radio Frequency Photomultiplier, Absolute calibration, Time of flight
\end{keyword}

\begin{abstract}
We propose a new method for absolute momentum calibration of magnetic spectrometers used in nuclear physics, using the time-of-flight (TOF) differences of pairs of particles with different masses. In cases where the flight path is not known, a calibration can be determined by using the TOF differences of two pair combinations  of three particles. A Cherenkov detector, read out by a radio frequency photomultiplier tube, is considered as the high-resolution and highly stable TOF detector. By means of Monte Carlo simulations it is demonstrated that the magnetic spectrometers at the MAMI electron-scattering facility can be calibrated absolutely with an accuracy $\delta p/p\leq 10^{-4}$, which will be crucial for high precision determination of hypernuclear masses. 
\end{abstract}}

\end{frontmatter}{}


\section{\label{sec:Introduction}Introduction}

The binding energy of the $\Lambda$ particle in the nuclear ground
state gives one of the basic pieces of information on the $\Lambda$-nucleus
interaction. This binding energy is defined by:

\begin{equation}
B_{\varLambda}=M_{core}+M_{\Lambda}-M_{HY}\label{eq:1}
\end{equation}

where $M_{core}$ is the mass of the nucleus that is left in the ground
state after the $\Lambda$ particle is removed, $M_{HY}$ is the mass
of the initial hypernucleus and $M_{\Lambda}$ is the mass of the
$\Lambda$. The binding energies $B_{\Lambda}$ have been measured
in emulsion for a wide range of light ($3\leq A\leq15$) hypernuclei
\cite{key-1,key-2,key-3}, exclusively from weak $\pi^{-}$ mesonic
decays. The binding energies of light hypernuclei provide the most
valuable experimental information to constrain various models of the
Y-N interaction. Table~\ref{tab:-separation-energies}, where the
numbers are taken from Ref.~\cite{key-4}, lists the results of the
$\Lambda$ separation energies obtained from ab initio theoretical
calculations using different Y-N interactions, along with the existing
experimental results. In addition to the quoted statistical errors,
the experiments also have systematic errors of about 0.04 MeV. It
is seen that precise experimental measurements of the binding energies
of light hypernuclei can discriminate between various models of Y-N
interactions. In particular, accurate measurements of the $\Lambda$
separation energies of light $\Lambda$-hypernuclei are a unique source
of information on charge symmetry breaking in the $\varLambda$-N
interaction and in $\Lambda$-hypernuclei \cite{key-5,key-6}. 

\begin{table}
{\scriptsize{}}%
\begin{tabular}{|c|c|c|c|c|c|c|}
\hline 
{\scriptsize{}Y-N} & {\scriptsize{}$B_{\Lambda}(_{\Lambda}^{3}H$)} & {\scriptsize{}$B_{\Lambda}(_{\Lambda}^{4}H$)} & {\scriptsize{}$B_{\Lambda}(_{\Lambda}^{3}H^{*}$)} & {\scriptsize{}$B_{\Lambda}(_{\Lambda}^{4}He$)} & {\scriptsize{}$B_{\Lambda}(_{\Lambda}^{4}He^{*}$)} & {\scriptsize{}$B_{\Lambda}(_{\Lambda}^{5}He$)}\tabularnewline
\hline 
\hline 
{\scriptsize{}SC97d(S)} & {\scriptsize{}0.01} & {\scriptsize{}1.67} & {\scriptsize{}1.20} & {\scriptsize{}1.62} & {\scriptsize{}1.17} & {\scriptsize{}3.17}\tabularnewline
\hline 
{\scriptsize{}SC97e(S)} & {\scriptsize{}0.10} & {\scriptsize{}2.06} & {\scriptsize{}0.92} & {\scriptsize{}2.02} & {\scriptsize{}0.90} & {\scriptsize{}2.75}\tabularnewline
\hline 
{\scriptsize{}SC97f(S)} & {\scriptsize{}0.18} & {\scriptsize{}2.16} & {\scriptsize{}0.63} & {\scriptsize{}2.11} & {\scriptsize{}0.62} & {\scriptsize{}2.10}\tabularnewline
\hline 
{\scriptsize{}SC89(S)} & {\scriptsize{}0.37} & {\scriptsize{}2.55} & {\scriptsize{}Unbound} & {\scriptsize{}2.47} & {\scriptsize{}Unbound} & {\scriptsize{}0.35}\tabularnewline
\hline 
{\scriptsize{}Experiment} & {\scriptsize{}$0.13\pm0.05$} & {\scriptsize{}$2.04\pm0.04$} & {\scriptsize{}$1.00\pm0.04$} & {\scriptsize{}$2.39\pm0.03$} & {\scriptsize{}$1.24\pm0.04$} & {\scriptsize{}$3.12\pm0.02$}\tabularnewline
\hline 
\end{tabular} 

\caption{$\Lambda$ \label{tab:-separation-energies}separation energies, $B_{\Lambda}$
given in units of MeV, of A = 3-5 $\Lambda$ hypernuclei for different
Y-N interaction models (see Ref. \cite{key-4}), compared to experiment.}

\end{table}

In 2007 the use of magnetic spectrometers to measure precisely the
momenta of pions from weak two-body decays of electroproduced hyperfragments
was proposed for Jefferson Lab \cite{key-7,key-8}. A similar experimental
program was started at the electron microtron in Mainz (MAMI) \cite{key-9},
where the first high resolution pion spectroscopy from decays of strange
systems was performed by electron scattering off a $\mathrm{^{9}Be}$
target \cite{key-10,key-11,key-12}. About 103 weak pionic decays
of hyperfragments and hyperons were observed. The pion momentum distribution
shows a monochromatic peak at $p\thickapprox133$~MeV/c, corresponding
to the unique signature for the two-body decay of hyper hydrogen $\mathrm{_{\Lambda}^{4}H\rightarrow\,_{\Lambda}^{4}He}+\pi^{-}$,
where the $_{\Lambda}^{4}\mathrm{H}$ stopped inside the target. Its
binding energy was determined to be $B_{\Lambda}=2.12\pm0.01$ (stat.)
$\pm0.09$ (sys.) MeV with respect to the $\mathrm{^{3}H}+\Lambda$
mass. 

We propose a new method for absolute calibration of magnetic spectrometers
by using the time-of-flight (TOF) differences of pairs of particles.
For example, by using the TOF difference of electrons and pions, a
magnetic spectrometer can be calibrated at momenta $\sim m_{\pi}c$
if the flight path is known. In cases where the flight path is not
known, a calibration can be determined by using the TOF differences
of two pair combinations of three particles, e.g. of electrons, pions
and electrons, kaons, or positrons, kaons and positrons, protons,
or positrons, protons and positrons, deuterons at momenta around $m_{\pi}c,$
$m_{K}c$, $m_{p}c$ respectively, where $m_{\pi},m_{K},m_{p}$ are
the masses of pions, kaons and protons. A new, ultra-high resolution
timing technique, based on Cherenkov radiation and the recently developed
Radio Frequency Photomultiplier Tube (RFPMT) \cite{key-13,key-14},
is considered. By means of Monte Carlo (MC) simulations it is demonstrated
that the magnetic spectrometers at MAMI can be calibrated absolutely
in the momentum range around 100 MeV/c with an error less than 0.01~MeV/c.
This would consequently decrease the systematic error of the binding
energy of hyper hydrogen to less than 0.007~MeV.

The decay pion experiment at MAMI is described in Sec.~\ref{sec:Hyperfragment-Electro}.
The method of absolute calibration of magnetic spectrometers using
TOF measurement of electrons and pions is presented in Sec.~\ref{sec:Absolute-Calibration}.
In Sec.~\ref{sec:The-Radio} and \ref{sec:The-RF-Deflector} the
operational principles of the RFPMT are described. Sec.~\ref{sec:An-Ultra-Precise}
is devoted to the ultra precise timing technique based on detection
of Cherenkov radiation using the RFPMT. Demonstration of the absolute
calibration of magnetic spectrometer SpekC at MAMI, by means of TOF
difference measurements of one or two pairs of particles, using MC
simulations, is presented in Sec.~\ref{sec:Abs-cal-pair} and Sec.~\ref{sec:Absolute-Calibration-1},
respectively. In the first case it is assumed that the flight path
is known; while in the second case TOF difference measurement of two
pair combinations of three particles determine the absolute values
of the flight path and momentum. Practical issues are discussed in
Sec.~\ref{sec:Practical-Issues}.

\section{\label{sec:Hyperfragment-Electro}The Hyperfragment Electroproduction
Experiment at MAMI}

\begin{figure}
\includegraphics[width=1\columnwidth]{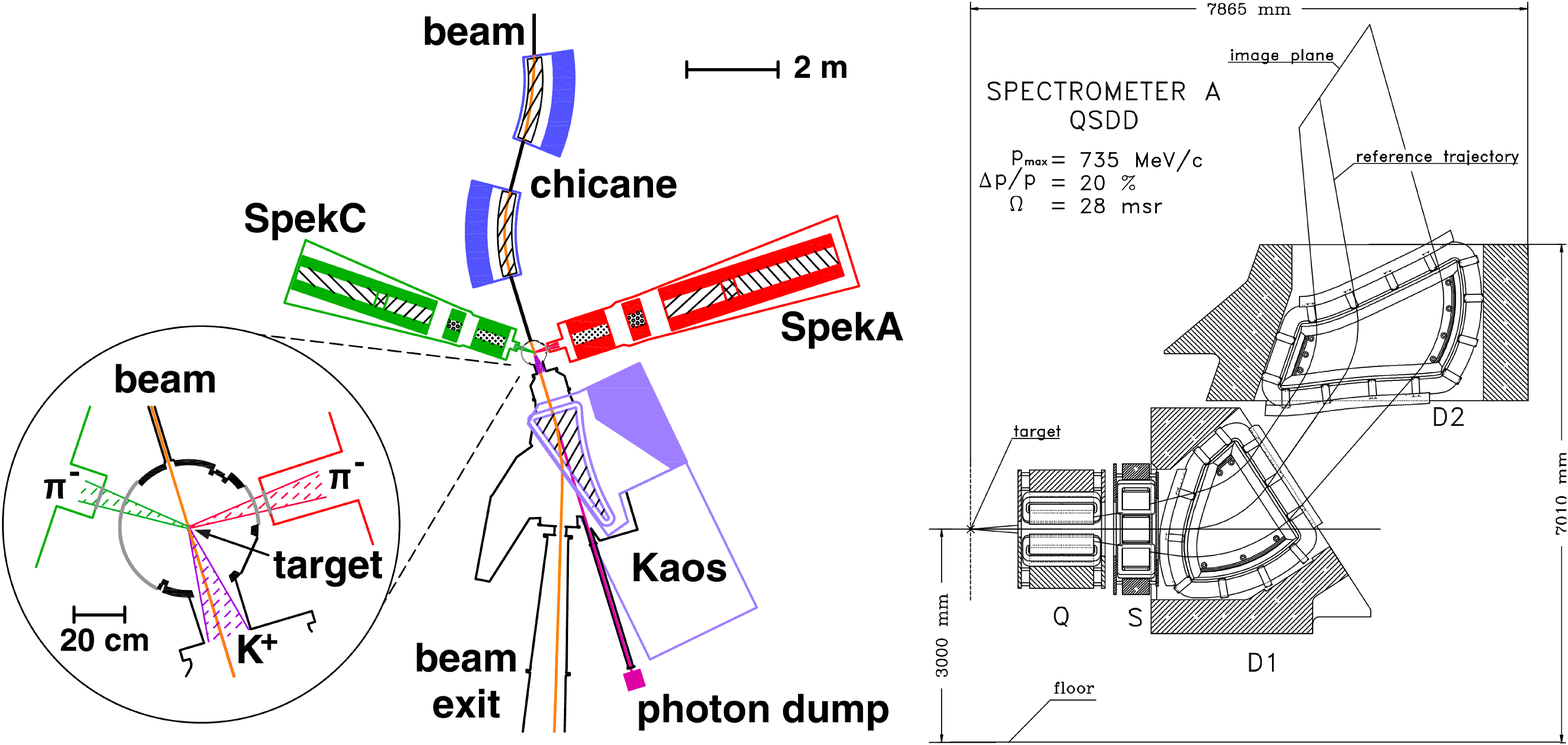}

\caption{\label{fig:Left:-floor-plan}Left: floor plan of electron beam-line
and magnetic spectrometers in the experimental hall at the Mainz Microtron
MAMI. Right: detail of the QSDD Spectrometer SpekA.}

\end{figure}

This experiment was carried out by the A1 Collaboration at the spectrometer
facility at MAMI \cite{key-15} (Fig.~\ref{fig:Left:-floor-plan}
left). A 1.508 GeV electron beam with a current of 20~$\mu$A was
incident on a 125~$\mu$m thick $^{9}Be$ target foil angled at 54
degrees with respect to the beam direction. Pions were detected with
two high-resolution spectrometers (SpekA and SpekC), each having a
quadrupole-sextupole-dipole-dipole, QSDD, configuration and a solid
angle of 28 msr (Fig.~\ref{fig:Left:-floor-plan} right). Vertical
drift chambers (VDCs), situated close to the image plane (Fig.~\ref{fig:Left:-floor-plan}
right) were used for tracking, scintillation detectors for triggering
and timing, and gas Cherenkov detectors for discrimination between
electrons and pions. The VDCs are capable of measuring a particle
track with effective position and angle resolutions of $\sigma_{x}=180\,\mu m$
and $\sigma_{\theta}=1$~mrad respectively. The spectrometers achieve
a relative momentum resolution of $\delta p/p=10^{-4}$ and were operated
at central momenta of 115 (SpekA) and 125 (SpekC) MeV/c with momentum
acceptances of $\varDelta p/p=20$\% (SpekA) and 25 \% (SpekC). The
tagging of kaons was performed by the Kaos spectrometer, positioned
at zero degrees with respect to the electron beam direction. The central
momentum was 924 MeV/c, covering a momentum range of $\varDelta p/p=50$\%
with a solid angle acceptance of $\Omega_{K}^{lab}=16$~msr.

\section{\label{sec:Absolute-Calibration}Absolute Calibration of Magnetic
Spectrometers}

Absolute calibration of the magnetic spectrometers at MAMI can be
realized by a TOF measurement of promptly produced pions and electrons
\cite{key-16}. Indeed, the TOF of pions or electrons of momentum
$p$ over a flight path $L$ is:

\begin{equation}
t_{\pi}=\frac{L}{\beta_{\pi}c}=\frac{L}{c}\left[1+\frac{m_{\pi}^{2}c^{2}}{p^{2}}\right]^{1/2}\label{eq:3}
\end{equation}

\begin{equation}
t_{e}=\frac{L}{\beta_{e}c}=\frac{L}{c}\left[1+\frac{m_{e}^{2}c^{2}}{p^{2}}\right]^{1/2}\label{eq:4}
\end{equation}

where $c$ is the speed of light and $\beta=v/c.$ From Eq.~\ref{eq:3},\ref{eq:4}
it follows that:

\begin{equation}
\frac{L}{c}=\left[\frac{t_{e}^{2}m_{\pi}^{2}-t_{\pi}^{2}m_{e}^{2}}{m_{\pi}^{2}-m_{e}^{2}}\right]^{1/2}\label{eq:5}
\end{equation}

\begin{equation}
p_{\pi}=\frac{L}{c}\frac{m_{\pi}c}{\left[t_{\pi}^{2}-(L/c)^{2}\right]^{1/2}}\label{eq:6}
\end{equation}

From Eq. \ref{eq:5},\ref{eq:6} $L/c$ and $p_{\pi}$ can be determined
uniquely by measuring $t_{\pi}$ and $t_{e}$. It is assumed that,
for a fixed configuration of the system, the flight path and the magnetic
rigidity of the spectrometer stay stable within a relative precision
better than $10^{-4}$. A similar technique has been established for
precise measurements of masses of exotic nuclei \cite{key-17}. However,
by using the RFPMT Cherenkov detector at MAMI we can determine absolute
calibrations by measuring only the TOF differences for single and
double pairs of particles. The details are described in Sec.~\ref{sec:Abs-cal-pair}
and Sec. \ref{sec:Absolute-Calibration-1}.

\section{\label{sec:The-Radio}The Radio Frequency Photomultiplier Tube}

\begin{figure}
\includegraphics[width=1\columnwidth]{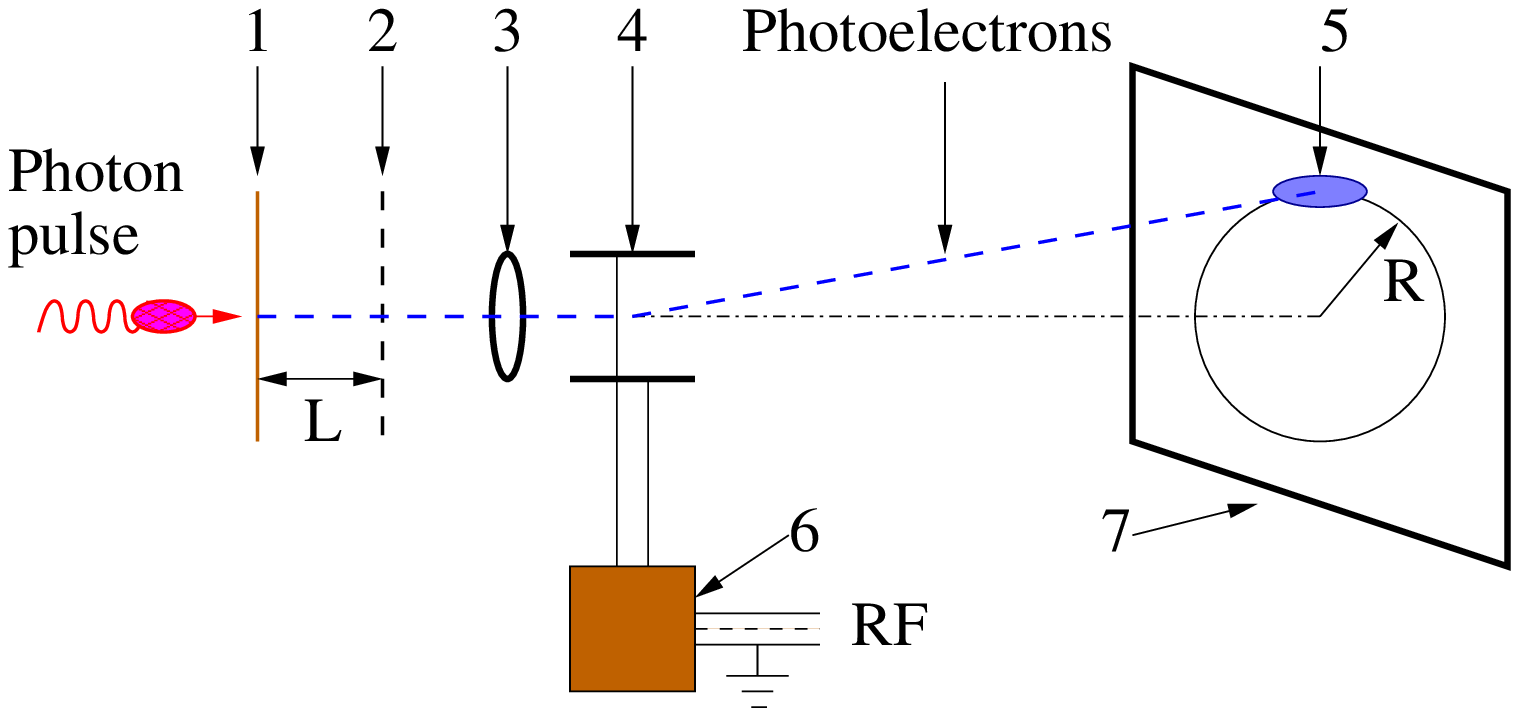}

\caption{\label{fig:Schem1-RFPMT}Schematic diagram of a small area photocathode
RFPMT: 1-photocathode, 2-electron transparent electrode, 3-electrostatic
lens, 4-electrodes of the RF deflector, 5-spot of PE on the PE detector,
6-$\lambda/4$ coaxial RF cavity, 7-PE detector.}

\end{figure}

A schematic diagram of a RFPMT with a small size cathode is given
in Fig.~\ref{fig:Schem1-RFPMT}. Incident photons strike the photocathode,
producing photoelectrons (PE) which are accelerated to 2.5 keV between
the photo cathode and an electron-transparent electrode. They are
then focused in an electrostatic lens and pass through the circular-sweep
RF deflection system. PE\textquoteright s passing through the RF deflector
form a circle on the screen of the PE detector, where the time structure
of the input photon signal is transformed into the spatial structure
of the electron image on a circle. 

\begin{figure}
\includegraphics[width=1\columnwidth]{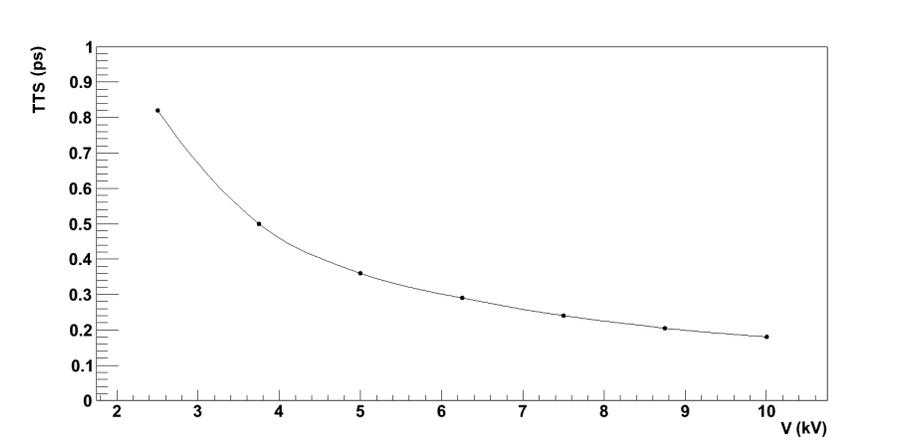}

\caption{\label{fig:Simulated-TTS}Simulated TTS vs applied accelerating voltage,
small size cathode RFPMT.}

\end{figure}

The detection of the RF analyzed PE's is accomplished with a position
sensitive (PS) PE detector. The time resolution for a single PE of
this RF timing technique, $\varDelta\tau_{RF}=\left(\varDelta\tau_{l}^{2}+\varDelta\tau_{d}^{2}\right)^{1/2}$,
is determined by the transit time spread (TTS), $\varDelta\tau_{l}$,
of PE's in the electron tube and the time resolution of the RF deflector,
$\varDelta\tau{}_{d}$. The transit time spreads were simulated by
means of SIMION 8 software \cite{key-18}. For an optimized tube geometry
the calculated TTS as a function of applied accelerating voltage is
shown in Fig.~\ref{fig:Simulated-TTS}. In the simulations, PE energies
were assumed to be distributed uniformly in the range 0-1 eV, while
their initial directions were taken to be isotropic. 

By definition the RF deflector time resolution is $\varDelta\tau_{d}=D/v$,
where $D$ is a convolution of the size of the PE beam spot on the
detector screen and the position resolution of the detector, and $v$=2$\pi R/T$
is the scanning speed. Here $R$ is the radius of the circle and $T$
is the period of the RF field. For a properly designed 1 GHz RFPMT
and PE detector with position resolution less than 0.1 mm, $\varDelta\tau_{d}\approx1$~ps. 

\begin{figure}
\includegraphics[width=1\columnwidth]{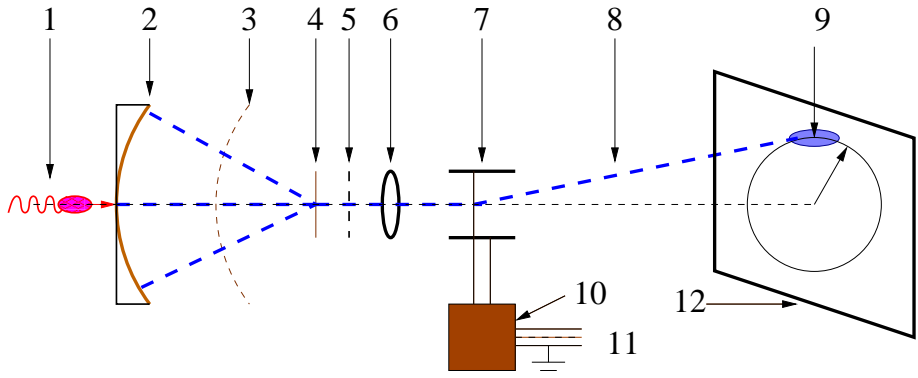}

\caption{\label{fig:schem2-RFPMT}The schematic layout of the RFPMT with an
extended size photocathode. 1-incident photon, 2-photo cathode, 3-electron
transparent electrode, 4-transmission dynode, 5-accelerating electrode,
6-electrostatic lens, 7-RF deflection electrodes, 8-RF deflected SE,
9-spot of SE on the PE detector, 10-RF coaxial cavity, 11-RF input,
12-PE detector. }
\end{figure}

\begin{figure}
\includegraphics[width=1\columnwidth]{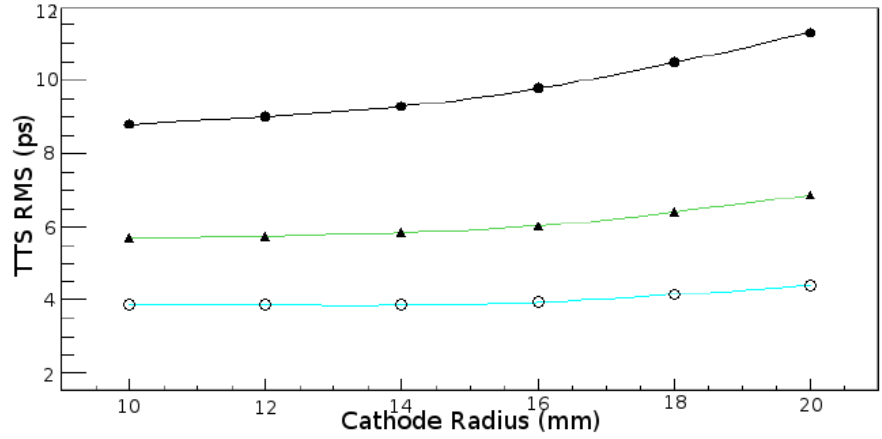}

\caption{\label{fig:TTS-Ext}Simulated TTS of the PE (at the cross over point
4 Fig.~\ref{fig:schem2-RFPMT}) vs cathode radius, for 3 applied
accelerating voltages between the cathode and accelerating electrode
(point 3 Fig. \ref{fig:schem2-RFPMT}). Filled circles 2.5 kV, triangles
5.0 kV, open circles 10 kV. Depicted from \cite{key-18}.}
\end{figure}

\begin{figure}
\includegraphics[width=1\columnwidth]{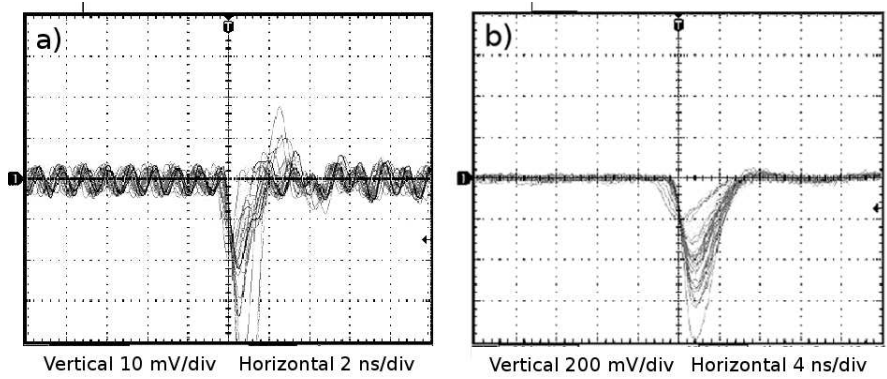}

\caption{\label{fig:RFPMT-signals}RFPMT signals collected on a 500~MHz digital
oscilloscope: a) directly from the anode, b) after a preamplifier.}

\end{figure}

A schematic diagram of the RFPMT with an extended photocathode is
given in Fig.~\ref{fig:schem2-RFPMT}. The primary photon pulse strikes
the photocathode and produces PE's. PE's are accelerated in the \textquotedblleft spherical-capacitor\textquotedblright{}
region and focused on the crossover where they pass through a transmission
dynode producing secondary electrons (SE) on both sides of the dynode.
Low energy SE's produced on the rear side of the transmission dynode
are accelerated by the electron transparent electrode and enter into
the electrostatic lens. SE's passing through the RF deflector are
deflected onto a circle on the screen of the PE detector and detected,
similar to the case of the small-size cathode. 

The TTS of PE's at the crossover, simulated by means of SIMION 8 software,
as a function of applied accelerating voltage between the cathode
and the accelerating electrode are shown in Fig.~\ref{fig:TTS-Ext}.
For an optimized large size photocathode RFPMT, the TTS can be as
low as $\sim5$~ps.

\section{\label{sec:The-RF-Deflector}The RF Deflector and Anode Readout Architecture}

The RF deflector consists of a pair of helical deflection electrodes
\cite{key-19}. These electrodes form a wire cavity with a quality
factor $Q\geq100$. The resonant frequency can be fixed at the desired
value by using a $\lambda/4$ coaxial cavity or an additional variable
capacitor. The sensitivity of the RF deflector at resonance frequencies
is about 0.1 rad/W$^{1/2}$ and a $\sim20$~V (peak to peak) RF sine
wave is sufficient to produce a scanning circle with a few cm radius
and line-width $D$ on the PE detector plane. This new RF deflector
can be operated in the 500-1000 MHz frequency range. 

We have tested the operational principles of a PE detector consisting
of a dual MCP chevron assembly, followed by an anode from which charge
is collected. The $\sim$ns rise time signal, generated by circularly
scanned 2.5 keV electrons incident on the dual MCP chevron assembly
and collected on the position sensitive resistive anode is shown in
Fig.~\ref{fig:RFPMT-signals}a. The signal after a preamplifier stage
is displayed in Fig.~\ref{fig:RFPMT-signals}b. The signal from the
anode (Fig.~\ref{fig:RFPMT-signals}a) consists of two parts: signals
generated by 2.5 keV electrons and pickup from the RF driving the
deflector. The single electron induced signals are an order of magnitude
larger than the RF induced pickup and they can be processed by regular
fast electronics. The few-ns integration time constant of the preamplifier
stage is sufficient to suppress the RF background almost entirely.

Two readout methods have been devised to locate the position on the
scanned circle: interpolation readout or pixel-by-pixel readout. Interpolation
readout, using a circular resistive anode, needs only two readout
channels and position is determined by applying charge-division or
delay-line time difference techniques. However it can only bear a
moderately high counting rate. For example with the dual MCP chevron
assembly the anticipated maximum rate is about 1 MHz. A pixel-by-pixel
readout anode will permit much higher counting rates \cite{key-20}
and pixel ASICs with 55 $\mu$m resolution are readily available.
A prototype resistive-anode RFPMT is under construction and will shortly
be tested at an electron accelerator facility.

\section{\label{sec:An-Ultra-Precise}An Ultra Precise Timing Technique based
on a Cherenkov Radiator and the RFPMT}

\begin{figure}
qq\includegraphics[width=1\columnwidth]{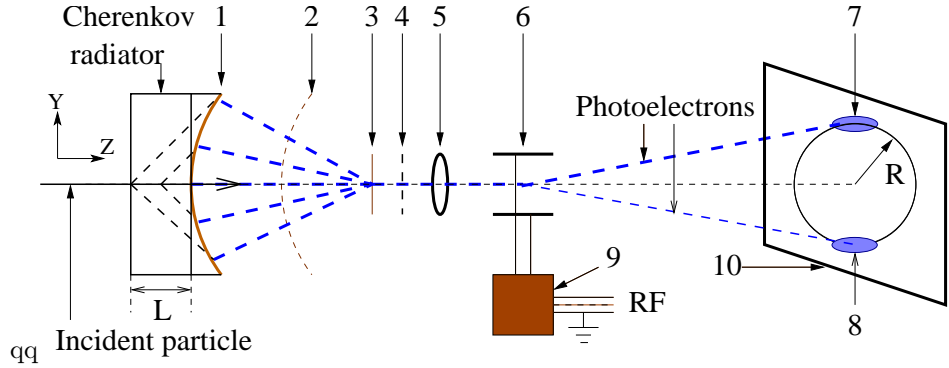}

\caption{\label{fig:schem3-RFPMT}A schematic layout of the Cherenkov TOF detector
with RFPMT. 1-photocathode, 2- electron transparent accelerating electrode,
3-transmission dynode, 4-accelerating electrode, 5-electrostatic lens,
6-RF deflection electrodes, 7-image of PE's from electron, 8-image
of PE's from pion, 9-RF coaxial cavity, 10-SE detector.}

\end{figure}

When a charged particle passes through a bar of transparent material,
Cherenkov photons are emitted in a cone defined by the Cherenkov angle
$\theta_{c}$, where $\cos\theta_{c}=1/n\beta$ and $n$ is the refractive
index. Cherenkov radiation is produced if the particle velocity $\beta>1/n$
and the flash duration of the Cherenkov radiation is $\leq1$~ps
\cite{key-21}. The paths of the Cherenkov photons in a radiator are
determined by $\theta_{c}$ (i.e. by the particle velocity) and the
azimuthal angle $\phi_{c}$ \cite{key-22}. These characteristics
of the Cherenkov radiation, in combination with a ps-resolution photon
detector, can produce an ultra high resolution timing detector. Here
we will consider a Cherenkov TOF detector based on the RFPMT and a
PbF$_{2}$ radiator ($n=1.82$). The following dominant factors have
been taken into account in the MC simulations \cite{key-13}: 
\begin{enumerate}
\item The time spread of Cherenkov radiation along the particle trajectory,
over the thickness of the radiator, where Cherenkov photons were emitted
uniformly along the particle track through the radiator.
\item The transit time spread of Cherenkov photons due to different trajectories
which, for individual photons, were determined according to $\theta_{c}$
and $\phi_{c}$.
\item The dependence of $n$ on the wavelength of the Cherenkov light, where
we take a mean value $n=1.82$ and assume a Gaussian distribution
for $n$ with $\sigma_{n}=0.008$.
\item The timing precision of the photon detector, where a Gaussian distribution
with $\sigma=10\,$ps has been used.
\end{enumerate}

\subsection{\label{sub:The-Cherenkov-TOF}The Cherenkov TOF detector in a head-on
geometry}

\begin{figure}
\includegraphics[width=1\columnwidth]{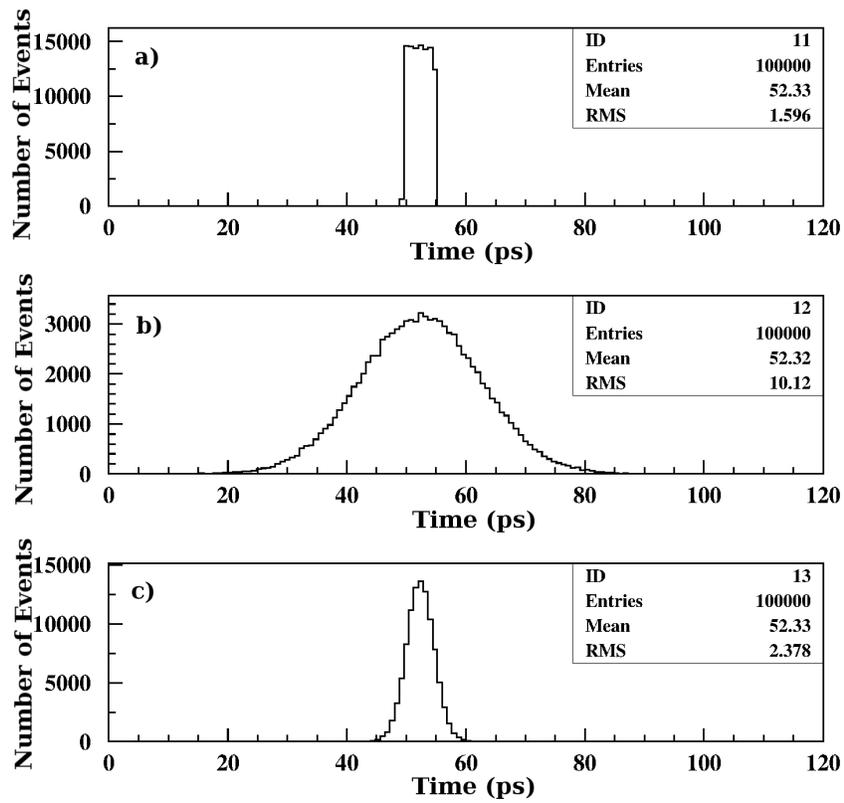}

\caption{\label{fig:MCsim1}MC simulated time distributions of (a) single Cherenkov
photons; (b) PE\textquoteright s for tracks of $p=133$~MeV/c pions;
(c) the mean time of 20 PE\textquoteright s. }
\end{figure}

\begin{figure}
\includegraphics[width=1\columnwidth]{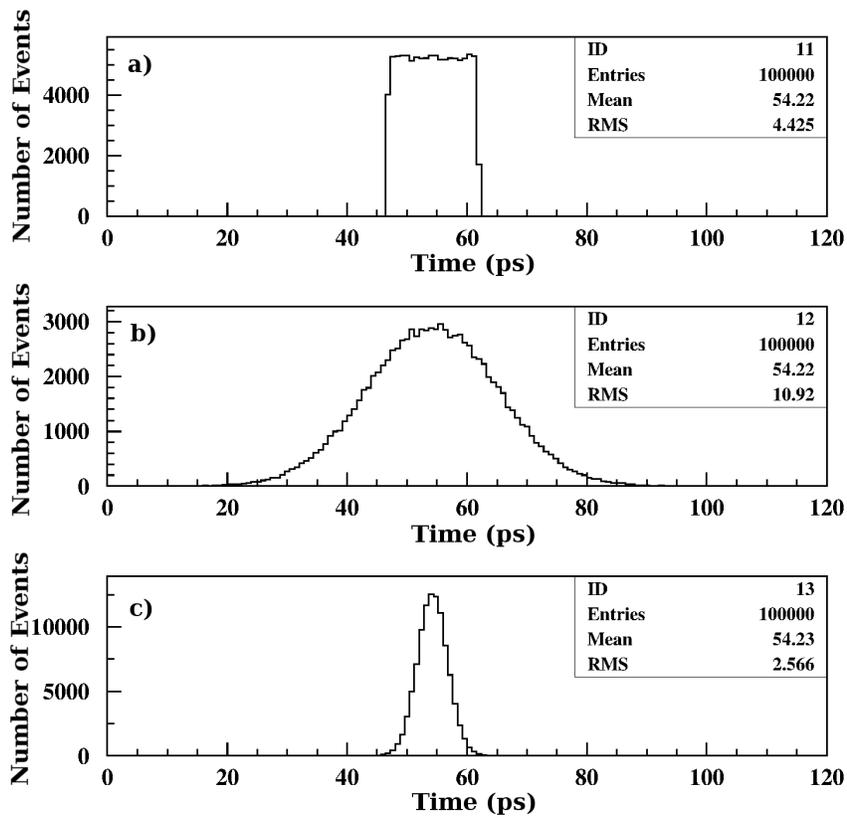}

\caption{\label{fig:MCsim2}MC simulated time distributions of (a) single Cherenkov
photons; (b) PE\textquoteright s for tracks of $p=133$~MeV/c electrons;
(c) the mean time of 20 PE\textquoteright s. }
\end{figure}

The Cherenkov TOF detector in a \textquotedblleft head-on\textquotedblright{}
geometry is shown schematically in Fig.~\ref{fig:schem3-RFPMT} The
incident particle produces Cherenkov photons in the radiator. These
photons produce PE's on the extended photocathode (1) which then pass
through the tube as in Sec.~\ref{sec:The-Radio}. The expected time
distribution of Cherenkov photons (a), single PE\textquoteright s
(b) and the mean of 20 PE\textquoteright s (c), for normally incident
133 MeV/c pions and electrons on a PbF$_{2}$ radiator (thickness
$L=0.2$~cm) for a RFPMT with 10 ps time resolution, are displayed
in Fig.~\ref{fig:MCsim1} and Fig.~\ref{fig:MCsim2}, respectively.
The time-zero, $T_{0}=40$~ps, is the time when a particle enters
the radiator. The delay time of PE's inside the RFPMT tube is assumed
constant and has not been considered in these simulations. These simulations
have demonstrated that such a Cherenkov detector can provide a time
resolution $\sim5$~ps FWHM. A similar result was obtained from a
simulation using the GEANT-4 software package.

\section{\label{sec:Abs-cal-pair}Absolute Calibration by a TOF Measurement
of a Pair of Particles}

\begin{figure}
\includegraphics[width=1\columnwidth]{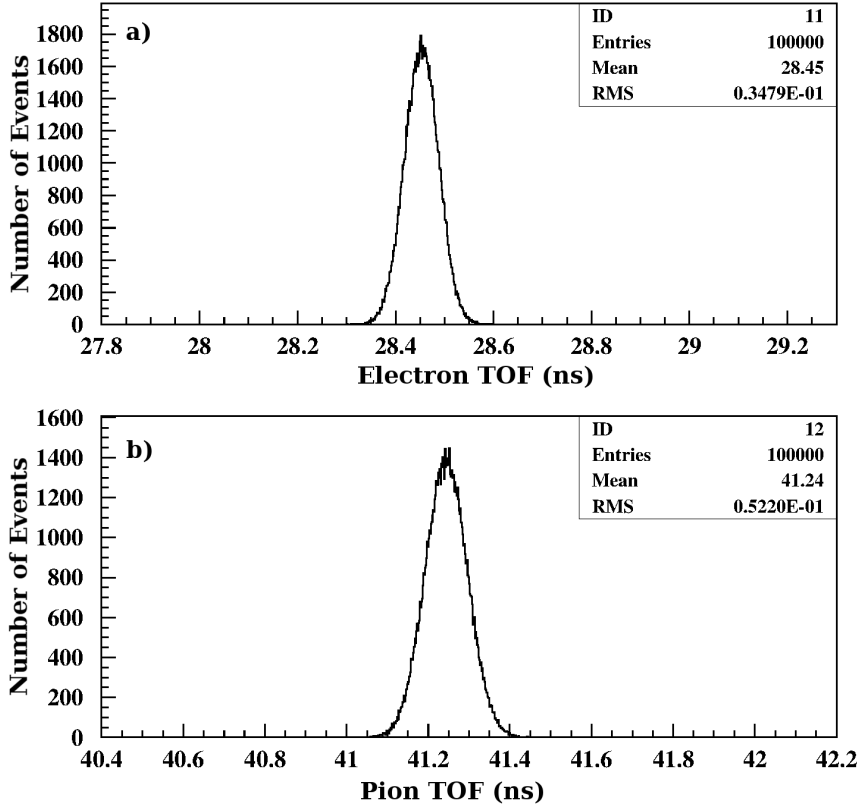}

\caption{\label{fig:MCsim3}The MC simulated distributions of $t_{e}$ (a)
and $t_{\pi}$ (b) for $L=853$~cm and $\sigma_{L}=1.0$~cm. The
mean values of these distributions are: $t_{e}^{av}=28453.2$~ps
and $t_{\pi}^{av}=41244.8$~ps.}

\end{figure}
We propose to use a Cherenkov detector in head on geometry for a TOF
measurement of pions and electrons at MAMI. The detector will be located
close to the focal plane of the spectrometer, with the RFPMT operated
synchronously with the electron bunches \cite{key-16,key-23} produced
by the MAMI accelerator. Both the bunch frequency (76.53~MHz) and
the RFPMT drive frequency (612.25~MHz) would be derived as sub harmonics
of the basic 2449~MHz operating frequency of the LINACs. In this
case PE's from Cherenkov radiation produced in the PbF$_{2}$ radiator,
by electrons and pions with the same momentum, will be located at
different places of the scanning circle of the RFPMT (Fig. \ref{fig:schem3-RFPMT}).
The flight times of pions $t_{\pi}^{i}$ and electrons $t_{e}^{j}$,
from the target to the Cherenkov radiator can be expressed in terms
of the period of the applied RF sinusoidal Voltage $T_{0}$:

\begin{equation}
t_{\pi}^{i}=N_{\pi}T_{0}+\Phi_{\pi}^{i}T_{0}\label{eq:7}
\end{equation}

\begin{equation}
t_{e}^{j}=N_{e}T_{0}+\Phi_{e}^{j}T_{0}\label{eq:8}
\end{equation}

where $N_{\pi}$, $N_{e}$ are integers and $\Phi_{\pi}^{i}$, $\Phi_{e}^{j}$
are the coordinates of the PE produced by pions and electrons on the
scanning circle of the RFPMT, relative to an arbitrary selected reference
(see Fig.~\ref{fig:schem3-RFPMT}). The distributions of $t_{\pi}$
and $t_{e}$ were obtained by means of MC simulations for MAMI SpekC.
The following factors have been taken into account: 
\begin{enumerate}
\item The timing accuracy of the RFPMT based Cherenkov detector, where a
Gaussian distribution with $\sigma=10$~ps has been used. 
\item The flight path of particles in the magnetic spectrometer: $L=853$~cm.
\item The flight path spread, where a Gaussian distributions with $\sigma_{L}=1.0$~cm
has been used.
\item The momentum distribution of electrons and pions, where we take $p=133$~MeV/c
and assume a Gaussian distribution for the momentum spread with $\sigma_{p}=0.1$~MeV/c. 
\end{enumerate}
The obtained distributions of $t_{\pi}$ and $t_{e}$ are shown in
Fig.~\ref{fig:MCsim3}. The histogram bin width is consistent with
the expected resolving power of the RFPMT-Cherenkov detector. 

The average times $t_{\pi}^{av}$, $t_{e}^{av}$ and their difference
can also be written: 

\begin{equation}
t_{\pi}^{av}=N_{\pi}T_{0}+\Phi_{\pi}^{av}T_{0}\label{eq:9}
\end{equation}

\begin{equation}
t_{e}^{av}=N_{e}T_{0}+\Phi_{e}^{av}T_{0}\label{eq:10}
\end{equation}

\begin{equation}
\varDelta T_{\pi e}=t_{\pi}^{av}-t_{e}^{av}=(N_{\pi}-N_{e})T_{0}+(\Phi_{\pi}^{av}-\Phi_{e}^{av})T_{0}\label{eq:11}
\end{equation}

The integers $N_{\pi}$ and $N_{e}$ were determined by using the
parameters of the SpekC given above. For example, if $T_{0}=1.96$~ns
we have $N_{\pi}=21$ and $N_{e}=14$. Therefore, taking these parameters,
$T_{\pi e}$ can be determined from Eq.~\ref{eq:11} by measuring
$\Phi_{\pi}^{av}-\Phi_{e}^{av}$. In this way the time delay in the
RFPMT is canceled, because it is the same for photo electrons from
pion and electron Cherenkov radiation. Rewriting Eq.~\ref{eq:5}
in the following form: 

\begin{equation}
\left[\frac{L}{c}\right]^{2}=\frac{t_{e}^{2}m_{\pi}^{2}-(t_{e}+\varDelta T_{\pi e})^{2}m_{e}^{2}}{m_{\pi}^{2}-m_{e}^{2}}\label{eq:12}
\end{equation}

\begin{equation}
t_{e}^{2}-\frac{2t_{e}\varDelta T_{\pi e}m_{e}^{2}}{m_{\pi}^{2}-m_{e}^{2}}-\frac{\varDelta T_{\pi e}^{2}m_{e}^{2}}{m_{\pi}^{2}-m_{e}^{2}}-\left[\frac{L}{c}\right]^{2}=0\label{eq:13}
\end{equation}

and solving Eq.~\ref{eq:13} for $t_{e}$ gives:

\begin{equation}
t_{e}=0.5\left[\frac{2\varDelta T_{\pi e}m_{e}^{2}}{m_{\pi}^{2}-m_{e}^{2}}\right]+\sqrt{D}\label{eq:14}
\end{equation}

where

\begin{equation}
D=4\varDelta T_{\pi e}^{2}\left[\frac{m_{e}^{2}}{m_{\pi}^{2}-m_{e}^{2}}\right]^{2}+4\left[\varDelta T_{\pi e}^{2}\frac{m_{e}^{2}}{m_{\pi}^{2}-m_{e}^{2}}+\left(\frac{L}{c}\right)^{2}\right]\label{eq:15}
\end{equation}

\begin{figure}
\includegraphics[width=1\columnwidth]{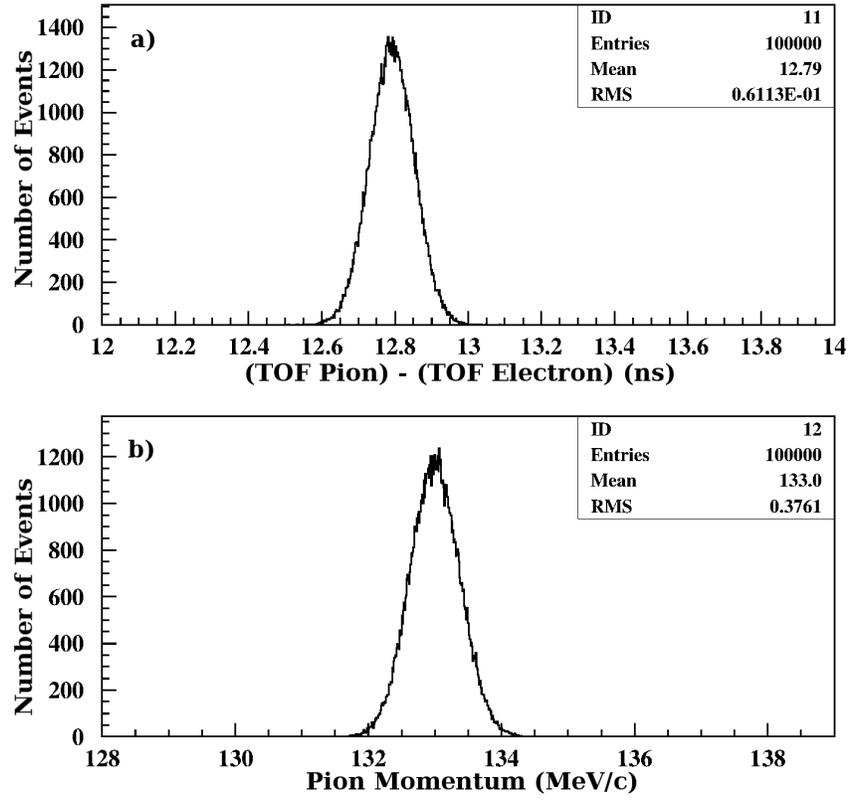}

\caption{\label{fig:MCsim4}The distributions of $t_{\pi}-t_{e}$ (a) and $p_{\pi}$(b)
for $L=853$~cm and $\sigma_{L}=1.0$~cm. The mean values of these
distributions are: 12791.4 ps and 133.00 MeV/c respectively.}

\end{figure}

The flight time for pions, $t_{\pi}=t_{e}+\varDelta T_{\pi e}$, and
consequently the absolute value of momentum $p$ can be determined
from Eq.~\ref{eq:6}. For example by using the mean values of the
distributions displayed in Fig.~\ref{fig:MCsim3} ($t_{e}^{av}=28453.2$~ps
and $t_{\pi}^{av}=41244.8$~ps) we obtain 132.99 MeV/c for the momentum
$p$. In principle we can also carry out this procedure by using the
$t_{\pi}$ and $t_{e}$ event-by-event for every individual pion and
electron pair, rather than taking their average values. The resulting
distributions of $t_{\pi}-t_{e}$ and $p$ are displayed in Fig.~\ref{fig:MCsim4}.
There are no significant differences in the values of $p$ obtained
using either average or event-by-event methods, but in the case of
an event-by-event analysis, any long term RFPMT time drifts are canceled
\cite{key-23}. 

\begin{table}
\begin{center}%
\begin{tabular}{|c|c|c|c|c|c|}
\hline 
$p_{in}$(MeV/c) & 132.0 & 132.5 & 133.0 & 133.5 & 134.0\tabularnewline
\hline 
$\varDelta T_{\pi e}$ (ps) & 12955.4 & 12873.1 & 12791.4 & 12710.5 & 12630.4\tabularnewline
\hline 
$p_{MC}$ (MeV/c) & 132.004 & 132.504 & 133.004 & 133.504 & 134.004\tabularnewline
\hline 
\end{tabular}\end{center}

\caption{\label{tab:The-results-MC}The results of MC simulations ($10^{5}$
events) for $\varDelta T_{\pi e}=t_{\pi}^{av}-t_{e}^{av}$ and $p_{MC}$,
with $L=853$~cm, $\sigma_{L}=1.0$~cm, for different values of
initial momentum $p_{in}$. }

\end{table}

The results of MC simulations for different values of initial momentum
are presented in Table~\ref{tab:The-results-MC}. From these simulations
it follows that a 1 MeV/c difference in momentum produces a difference
in $\varDelta T_{\pi e}\thickapprox160$~ps over a flight path $L=853$~cm.
Therefore $\varDelta T_{\pi e}$ must be measured to a precision of
about 1.6~ps to give an absolute calibration of SpekC to better than
10~keV/c, which is achievable with the RFPMT. However the absolute
value of $L$ would also require to be known to a precision of 853~cm
$\times$ 1.6~ps/12791.4~ps = 1.07~mm, which would be very difficult
in practice.

\section{\label{sec:Absolute-Calibration-1}Absolute Calibration by a TOF
Measurement of a Triplet of Particles}

\begin{figure}
\includegraphics[width=1\columnwidth]{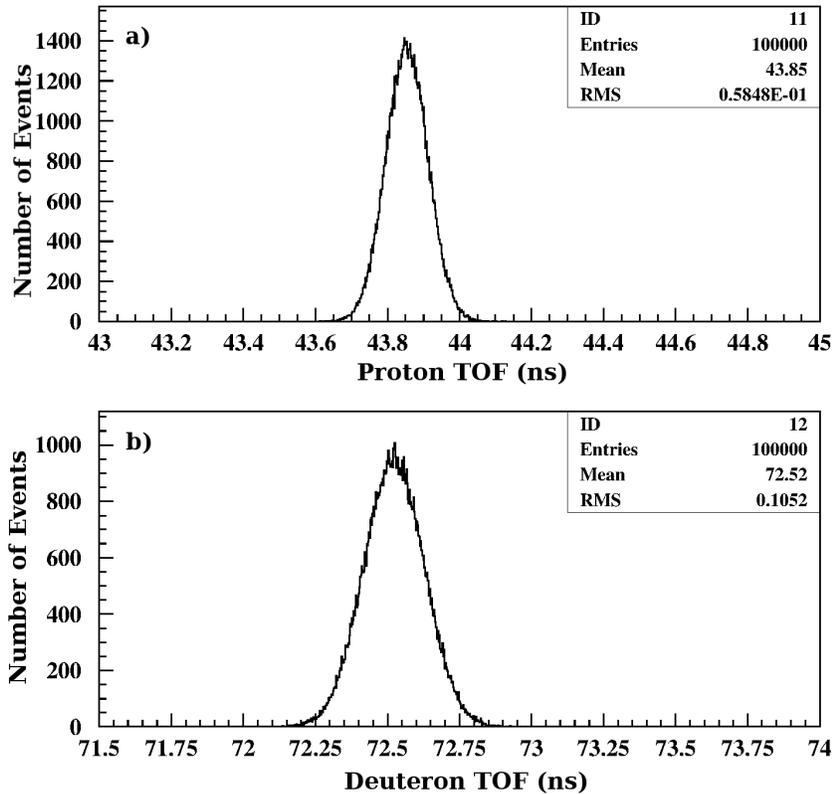}

\caption{\label{fig:MCsim5}The MC simulated distributions of $t_{p}$ (a)
and $t_{d}$ (b) for $L=853$~cm, $\sigma_{L}=1.0$~cm and $p=800$~MeV/c.
The mean values of these distributions are: $t_{p}^{av}=43851\,$ps
and $t_{d}^{av}=72522$~ps.}

\end{figure}

\begin{figure}
\includegraphics[width=1\columnwidth]{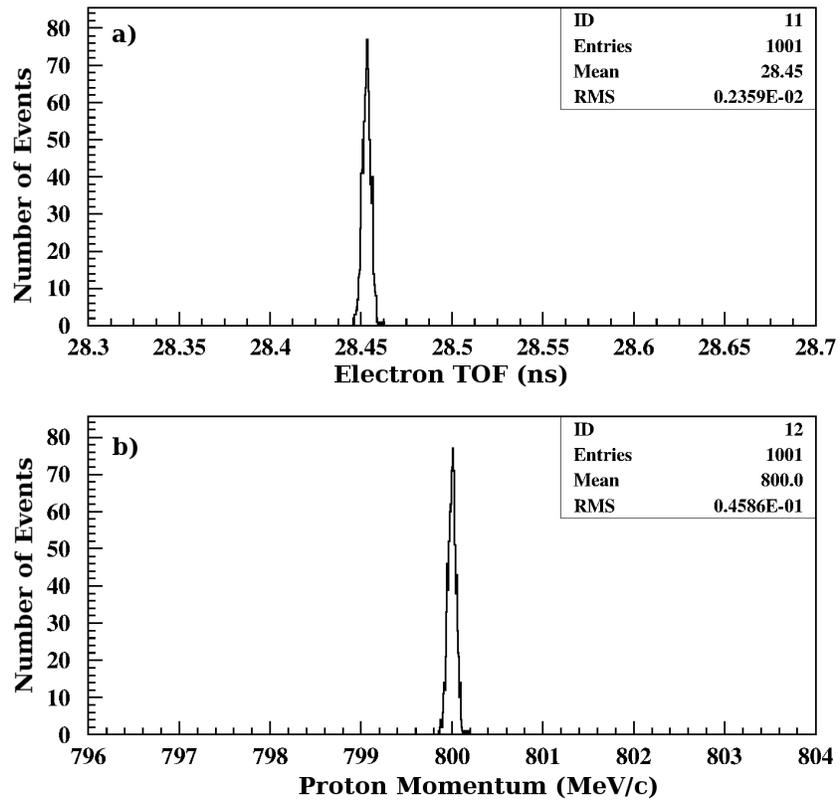}

\caption{\label{fig:MCsim6}The MC simulated distributions of the calibrated
values of $t_{e}$ (a) and $p_{p}$ (b). The mean values of these
distributions are: 28453 ps and 800.00 MeV/c.}

\end{figure}

The flight path uncertainty can be avoided if, in addition to the
electron and pion, a third particle, e.g. a kaon, is detected. An
absolute calibration of a magnetic spectrometer can be determined
by measurement of differences in TOF of the two pair combinations
e.g. pions and electrons, $\varDelta T_{\pi e}=t_{\pi}-t_{e}$, and
kaons and electrons, $\varDelta T_{Ke}=t_{k}-t_{e}$. Rewriting Eq.~\ref{eq:5}: 

\begin{equation}
\left[\frac{L}{c}\right]^{2}=\frac{t_{e}^{2}m_{K}^{2}-(t_{e}+\varDelta T_{Ke})^{2}m_{e}^{2}}{m_{K}^{2}-m_{e}^{2}}\label{eq:16}
\end{equation}

and deriving the following from Eq.~\ref{eq:6}

\begin{equation}
\left[\frac{m_{\pi}}{m_{e}}\right]^{2}=\frac{(t_{e}+\varDelta T_{\pi e})^{2}-(L/c)^{2}}{t_{e}^{2}-(L/c)^{2}}\label{eq:17}
\end{equation}

it follows that $t_{e}$ can be determined uniquely:

\begin{equation}
t_{e}=\frac{\varDelta T_{\pi e}^{2}(m_{K}^{2}-m_{e}^{2})+\varDelta T_{Ke}^{2}m_{e}^{2}-\varDelta T_{Ke}^{2}m_{\pi}^{2}}{2\left(\varDelta T_{Ke}m_{\pi}^{2}-\varDelta T_{\pi e}m_{K}^{2}+\varDelta T_{\pi e}m_{e}^{2}-\varDelta T_{Ke}m_{e}^{2}\right)}\label{eq:18}
\end{equation}

In this concept $t_{e}$ and, consequently $L$, $t_{\pi}$, $t_{K}$
and the absolute momentum $p$, are determined by $\varDelta T_{\pi e}$,
$\varDelta T_{Ke}$ and the masses of electrons, pions and kaons.
In principle other combination of particles can be used, e.g. positrons,
kaons and protons or positrons, protons and deuterons. Each combination
of particles is effective in a given momentum range, where the momenta
of at least 2 particles from three are non-relativistic. On the other
hand they have to have enough velocity to produce prompt Cherenkov
radiation. The effective momentum intervals are $\leq200$~MeV/c
for electrons, pions and kaons; $\leq500$~MeV/c for positrons, kaons
and protons; and $\leq1000$~MeV/c for positrons, protons and deuterons. 

Since kaons in the previously mentioned momentum range will mainly
decay over a 8.53 m flight path it may not be practical to use them.
Alternatively one can use positrons, protons and deuterons to calibrate
the magnetic spectrometer in the 800 - 1000~MeV/c momentum range
and determine the central flight path $L$. TOF distributions of protons
and deuterons, obtained from MC simulations for 800~MeV/c are shown
in Fig.~\ref{fig:MCsim5}. The average values of these distributions
were used to determine their differences and calculate the absolute
value of $t_{e}$ by using Eq.~\ref{eq:18}, where $\varDelta T_{\pi e},\:\varDelta T_{Ke},\:m_{\pi},\:m_{K}$
are replaced by $\varDelta T_{pe},\:\varDelta T_{de},\:m_{p},\:m_{d}$. 

The absolute value of $t_{e}$ determines the absolute calibration
of $L,\:t_{p},\:t_{d}$ and $p$ and the distributions of $t_{e}$
and $p$ obtained in this way are displayed in Fig.~\ref{fig:MCsim6}.
In this simulation a synthetic rutile crystal (Titanium dioxide, TiO$_{2}$,
$n=2.6$), was considered as a radiator \cite{key-24} in order to
extend Cherenkov sensitivity to lower velocities. The value of $L$
determined in this way can then be used for absolute calibration of
a magnetic spectrometer in the momentum range around 100 MeV/c by
using the TOF difference of pions and electrons as described in Sec.~\ref{sec:Abs-cal-pair}.
It is assumed that, for a fixed setup, the flight path length and
other parameters of the spectrometer stay stable within a precision
better than $10^{-4}$.

It is worth mentioning that Cherenkov radiation has already been registered
by a circular-scan streak camera in Synchroscan mode \cite{key-25}.
Synchroscan operation of streak cameras is a regular timing technique
for particle bunches at accelerators and the phase stability or time
drift of the technique is 1-2 ps over periods of hours \cite{key-26}.

\section{\label{sec:Practical-Issues}Practical Issues}

As an example we consider the absolute calibration of SpekC at momenta
around 133 MeV/c, which is close to the decay pion momentum of hyper
hydrogen $_{\Lambda}^{4}H$. If the flight path $L$ from the target
to the Cherenkov detector is known, the calibration can be determined
by TOF difference of electrons and pions. If the flight path from
target to Cherenkov detector is not known we need three particles.
As mentioned previously it may not be practical to use kaons due to
the relatively long flight path of 8.53 m. Therefore we propose to
use positrons, protons and deuterons to make a calibration at 800
MeV/c, which yields a value of $L$ with a precision better than 1~mm.
Using this calibrated value of $L$, and the TOF difference of pions
and electrons, the magnetic spectrometer can be calibrated at a momentum
range around 100 MeV/c, which is necessary for the decay pion experiment. 

The Cherenkov detector will be mounted after the spectrometer's drift
chambers which sit close to the image plane of the spectrometer (see
Fig.~\ref{fig:Left:-floor-plan}). Recently at MAMI the so-called
single pulse operation, with few ps electron bunches every $13.07$~ns,
was implemented and tested successfully up to average beam currents
of 40~$\mu$A. The bunch separation of $13.07$~ns follows from
the use of a laser with a repetition frequency locked to the 32nd
sub harmonic frequency of the MAMI standard microwave frequency of
2.449 GHz, that is 76.53 MHz. We can use this electron beam, but operate
the RFPMT with a frequency locked to the 4th sub harmonic frequency
of 2.449 GHz, which is 612.25 MHz. It is assumed that, by means of
the tracking detectors, particles with well defined momentum $\varDelta p/p\sim10^{-3}$
can be selected, leading to a flight path spread $\varDelta L\sim10$~mm. 

The Cherenkov detector, readout by the RFPMT, will provide $\sim$ns
rise time electrical signals. These signals can be used, in a regular
timing technique, to determine relative time differences between electrons
and pions (or other particles) in SpekC and positrons/electrons in
Kaos, with a precision of less than 1 ns \cite{key-11}. This information
will be used for identification of electrons and pions (or other particles)
in SpekC. 

Meanwhile charge comparison of the two signals from the RFPMT anode
can be used to determine the position of PE's on the scanning circle
with a precision of about 50~$\mu$m (see Fig.~\ref{fig:schem3-RFPMT}),
equivalent to timing the PE's with $\sim$1~ps precision, which would
determine precisely the TOF differences of pions and electrons or
any other pair of particles. This in turn is would determine the calibration
of the spectrometer.

\section{\label{sec:Conclusions}Conclusions}

A new method for absolute momentum calibration of magnetic spectrometers
employed in nuclear physics, using the time-of-flight (TOF) difference
of pairs of particles is proposed. The situation of electrons and
pions, for a known flight path, has been simulated at momenta around
$m_{\pi}c$, where $m_{\pi}$ is the pion mass. Cases where the flight
path is not known have been simulated by using the TOF differences
of two pair combinations of three particles, in this case positrons,
protons and positrons, deuterons at momenta close to $m_{p}c$, where
$m_{p}$ is the proton mass. This yields a high-precision value of
the flight path, which can then be fed back to the electron-pion case
at lower momentum. A Cherenkov detector, read out by a radio frequency
photomultiplier tube, has been simulated as the high-resolution and
highly stable TOF detector. The Monte Carlo simulations predict that
the technique, with the RFPMT operating at RF frequencies 500-1000
MHz, has a $\sim$10 ps resolution for single photons, and is capable
of achieving 1 ps stability levels over periods of hours. The calculations
predict that the magnetic spectrometers at the MAMI electron-scattering
facility can be calibrated absolutely at momenta around 100~MeV/c
with an accuracy $\delta p/p\sim10^{-4}$, which will be crucial for
precise determination of the binding energies of light hypernuclear
systems.

\section*{Acknowledgments}

This work was supported by the RA MES State Committee of Science,
within the framework of research project 15T-2B206 and by the UK Science
and Technology Facilities Council (Grant nos. STFC 57071/1 and STFC
50727/1).

\end{document}